\documentclass[12pt,preprint]{aastex}

\usepackage{epsfig}
\usepackage{amsmath,amssymb}
\usepackage[all]{xy}
\usepackage{graphicx}
\usepackage{natbib}

\usepackage{epstopdf}
\newcommand{\beq}{\begin{equation}}
\newcommand{\eeq}{\end{equation}}

\newcommand{\be}{\begin{equation}}
\newcommand{\ee}{\end{equation}}
\begin{document}

\title{A Cluster of Ultrahigh Energy Cosmic Rays }
\author{Glennys R. Farrar}
\affil{Center for Cosmology and Particle Physics, 
Department of Physics,New York University, 
New York, NY 10003, USA; gf25@nyu.edu}


\begin{abstract}
Five ultrahigh energy cosmic rays in the combined AGASA and HiRes stereo data are analyzed to test whether they come from a single source.  The quad above 37 EeV in the 94-event high energy dataset can be analyzed without considering magnetic dispersion.  The probability that it is a chance association is $10^{-3}$.  Assuming the source is continuous, the random magnetic deflections these UHECRs accumulated en route from their source can be used to estimate  $\sqrt{\langle B^2 \lambda \rangle D} \approx 1$ nG-Mpc.  A quintuplet including a HiRes event between 10 and 30 EeV is well fit with this value.  Galactic magnetic deflection appears to be smaller than in some models.
\end{abstract}


\section{Introduction}
The combined ultrahigh energy cosmic ray data from the AGASA and stereo HiRes detectors was recently studied using the maximum likelihood method by the HiRes collaboration and this author\citep{HRGF}. It was noted that a 37.6 EeV HiRes event and the previously-reported triplet from AGASA\citep{AGASAupdate} could have a common origin: the  distribution of their arrival directions is consistent with a pointlike source given the measurement errors of the events.   The frequency in random simulations of an equally strong clustering signal being observed by chance in a dataset of this size (57 AGASA events above 40 EeV and 37 HiRes events above 30 EeV) was reported to be about 0.5\% using the likelihood ratio alone.  In this Letter I point out that a more informative criterion for the strength of the cluster gives a chance probability of $10^{-3}$, and that the uncertainty or bias in the estimate due to definition of the dataset is minor.  Thus the evidence that these events come from a single source is quite strong, and justifies investigating the implications of the cluster assuming it is real.  Here, I concentrate on inferences which can be made about the magnetic field in the direction of the source.  I also investigate a nearby event in the 10-30 EeV HiRes data \citep{HRclus04}. It gives an excellent fit to coming from the same source, but given the size of the low energy dataset has about a 1 in 6 chance of being a random association.

Empirically, the four higher energy events have arrival directions so close that they can be analyzed without including magnetic deflection \citep{HRGF}.  Initially without magnetic deflection, I explain the method and estimate the chance probability for this cluster.   Next, magnetic deflection is added to the analysis and the observed arrival directions are fit to the hypothesis that they are spread not only by measurement error but also by dispersion due to random magnetic fields between source and detector.  Possible net deflection due to the regular local galactic magnetic field  is also considered. 

\section{Maximum Likelihood Method} 
If the magnetic dispersion along the line-of-sight to a source of UHECRs is small and the source is strong enough, the maximum likelihood method (ML) gives an important improvement over previous methods of studying clustering.  It incorporates the resolution of individual events and therefore avoids arbitrariness in the definition of a cluster, it allows data from different experiments to be combined, and it is the best way to estimate the direction of the source. The application of the maximum likelihood method to this problem is described in \citep{HRGF}.  One postulates the existence of a point source located at $\vec{\theta}_{s}$ contributing $n_{s}$ events to the dataset which has a total of $N$ cosmic ray events, the remaining $N - n_{s}$ events being random, isotropically distributed background events.  If the $i$th event is a source event, then its true arrival direction is $\vec{\theta}_{s}$ and the probability density for observing it in some direction $\vec{\theta}$ is denoted by $Q_{i}(\vec{\theta},\vec{\theta}_{s})$, which depends on the angular errors of the event and the angular displacement between $\vec{\theta}_{s}$ and $\vec{\theta}$. On the other hand, if the $i$th event is a background event, then the probability density for finding it at a location $\vec{\theta}$ is given by $R_{i}(\vec{\theta})$, the relative  exposure of the detector to an isotropic background of cosmic rays. Each of these functions is normalized over all directions $\vec{\theta}$ in the sky. Assuming $R_i$ is small, $N$ is large, and $N \gg n_s$, the partial probability distribution of a given arrival direction $\vec{\theta}$ for the $i$th event is simply
\begin{equation}
P_i(\vec{\theta}) = \frac{n_{s}}{N}\,Q_i(\vec{\theta},\vec{\theta}_{s})
                + \frac{N-n_{s}}{N}\,R_i(\vec{\theta})~~,
\end{equation}
and the likelihood for the dataset of $N$ events is $
{\mathcal L}(n_{s}) \equiv \prod_{i=1}^{N} P_{i}(\vec{\theta}_{i})$.
The best estimate for the source position $\vec{\theta}_{s}$ is the value which maximizes $\mathcal{L}$.  It is convenient to work with the logarithm of the (dimensionless) likelihood ratio, $ln \mathcal{R}$:
\begin{equation} \label{lnR}
{ln \mathcal R}(n_{s})  \equiv ln  \frac{{\mathcal L}(n_{s})}
        {{\mathcal L}(0)}
      = \sum_{i=1}^{N}~ ln
        \left\{\frac{n_{s}}{N}
           \left(\frac{Q_{i}(\vec{\theta}_{i},\vec{\theta}_{s})}
                      {R_{i}(\vec{\theta}_{i})}
           -1\right)
        +1\right\}  ~~,
\end{equation}
where ${\mathcal L}(0)$ is the likelihood function for $n_{s}=0$.  Below, $ln \mathcal R$ and $n_s$ denote the values of these quantities at the location and $n_s$ value which maximizes $ln \mathcal R$, unless otherwise specified.  

In the simplifying case of Gaussian resolution with a single variance $\sigma$, uniform exposure density $R$, average inter-event spacing large compared to $\sigma$, and $N \gg n_s \gg 1$, one can recast (\ref{lnR}) in the form
\begin{equation}
\label{lnRapprox1}
ln \mathcal{ R} = - \frac{1}{2} \chi^2   - ln  P_{\epsilon, n_s,N} 
\end{equation}
where $\chi^2 \equiv \sum_{source} \frac{\Delta \theta_i^2}{ \sigma^2}$, minimized with respect to source location, and 
 \begin{equation}
\label{chanceprob}
P_{\epsilon, n_s,N} = \frac{N!}{n_s! (N-n_s)!} \epsilon^{n_s} \approx \left( \frac{N \epsilon}{n_s} \right)^{n_s} 
\end{equation}
is the chance probability to get $n_s$ hits in $N$ tries, in a region whose fractional size is $\epsilon \equiv 2 \pi \sigma^2 R$.

Note that ${\mathcal L}$ is a probability {\it density} and not a probability, and $\mathcal R$ is not a ratio of probabilities.  To obtain the ratio of probabilities that the given observation is a signal versus being background, one must specify precisely what condition to impose defining the signal.  For instance, one could define the signal probability to be the fraction of times a source of $n_s$ events has a $\chi^2 \ge$ the one observed, and the background probability as the fraction of times a random distribution has $n_s$ events with $\chi^2 \le$ that observed.  Or, one could integrate ${\mathcal L}(\vec{\theta})$ over a region of $\vec{\theta}$ around the maximum $\vec{\theta}_s$, subject to some condition on the desired signal strength. 

Directly computing the signal-to-chance probability ratio by integrating ${\mathcal L}(n_s)$ and ${\mathcal L}(0)$ is cumbersome and an alternative approach is commonly used in experimental analyses in which many synthetic datasets are created (here, by isotropizing the observed events weighted according to the exposure of the detectors), and the number of cases having $ln \mathcal R \ge ln \mathcal R_{\rm obs}$ is tabulated.   However this does not in general provide an accurate estimate of the ratio of signal to background probabilities, because, as can be understood from equations (\ref{lnRapprox1}) and (\ref{chanceprob}),  since $\epsilon N$ is small, $ln \mathcal R$ can be large either because $n_s$ is large, or because $\chi^2$ is small.  The mean value of $\chi^2$ for a genuine source is equal to the number of degrees of freedom, and a low value of $\chi^2$ is {\it not} an indicator of a ``better" source.  Thus a better estimate of the chance probability of a given observed cluster is obtained by counting the number of cases in random trials in which both $ln \mathcal R \ge ln \mathcal R_{\rm obs}$ and $n_s \ge n_{s*}$.  For this purpose,  the simplest choice is $n_{s*} = n_{s, \rm obs}$, and that is used below.  Another option which Auger might wish to adopt is to choose (by simulation in advance of the analysis) values of $ln \mathcal R_*$ and $n_{s*}$ that maximize rejection for purely random datasets while keeping, say,  95\% of true clusters embedded in random background.  This procedure gives a minimum value of $n_{s*}$ which can be probed with a given density dataset, for a specified degree of background rejection. 

\section{Data and Simulations}
For AGASA, the relative exposure is quite uniform, with $R = 0.18~ {\rm sr}^{-1} = 5.5 \times 10^{-5} {\rm deg}^{-2}$.   The HiRes relative exposure is similar but less uniform, as shown in Fig. 1 of \citet{HRGF}.  In the region of the quad, $R_{HR} = 0.2~ {\rm sr}^{-1}$.  The resolution of stereo HiRes events is, with few exceptions, well-approximated as a symmetric, 2-d Gaussian with $\sigma_{HR} = 0.4^\circ$ \citep{HRGF}.   A definitive study of the clustering of the data must await publication of the full AGASA catalog, including individual angular errors, but we can make a preliminary analysis, as follows.  Individual AGASA events can be described by a symmetric 2-d Gaussian (M. Teshima, private communication) whose resolution varies with energy and depends on whether the event is well-contained or not.   A higher energy event can be reconstructed better because more detectors are hit and the statistical fluctuations within the detectors are lower.  Only detectors near the shower core are used in the angular reconstruction, so events whose core is at least 1 km inside the array (``well-contained" events) are better measured than those at the perimeter (merely ``inside").  60\% of the events above 40 EeV are ``well-contained" and we denote their resolution by $\sigma_g(E)$, while 40\% are merely ``inside" with resolution denoted $\sigma_b(E)$.  Fig. 1 of  \citet{AGASAclusters}, shows the opening angles containing 68\% and 90\% of events as a function of energy up to 100 EeV, allowing one to infer (last digit not significant):
\begin{eqnarray} \label{AGsigs}
\sigma_{g}(E) & = & 2.13^{\circ} - 0.80^{\circ} 
                   \log_{10}(E_{\mathrm{EeV}})~~~ \\
\sigma_{b}(E) & = & 4.49^{\circ} - 1.56^{\circ} 
                   \log_{10}(E_{\mathrm{EeV}})~.
\end{eqnarray}
The three events in the AGASA triplet are all well-contained (M. Teshima, private communication), so the analysis below uses $\sigma_g$ for those events, i.e., $0.62^\circ,~ 0.74^\circ$ and $0.75^\circ$ for the 77.6, 55, and 53.5 EeV events respectively.  In simulations, errors on other AGASA events are assigned at random between $\sigma_g(E)$ and $\sigma_b(E)$ in the 60\% - 40\% proportion and all HiRes events have $\sigma = 0.4^\circ$. 

There are 251 HiRes events in the fiducial region, $-10^\circ \le {\rm dec} \le 80^\circ$; their arrival directions can be taken from the skymap in \citet{HRclus04}.  In addition to the 37.6 EeV event appearing in the dataset above 30 EeV, one lower energy event is near the quad, at \{171.7$^\circ$ , 57.8$^\circ$\}.  Its energy will be denoted $E_5$.  Using an $E^{-2.7}$ spectrum, the average energy of events in the 10-30 EeV range is 15.4 EeV, so for definiteness $E_5 = 15$ EeV is adopted below.  Where relevant, the lack of knowledge of $E_5$ is taken into account.   Note that the energies of several events reported as above 40 EeV in \cite{AGclus96} were revised downward; here, as in \cite{HRGF}, the specifics of the AGASA events are taken from \cite{AGASAupdate}. 

There are three technical differences between the simulations here and in \citet{HRGF}. First, to find the most likely source location,  \citet{HRGF} scanned over a grid of points with about $0.1^\circ$ separation and took the position with the largest value of $ ln\mathcal{R}$.  Here, the best source location is roughly determined by a grid-based scan, then the exact point in $n_s$, RA, and dec which maximizes $ ln\mathcal{R}$ is found.  Second, \citet{HRGF} took the AGASA resolution to be a sum of two Gaussians, while here individual events are given Gaussian errors of different sizes, to more accurately reflect the actual conditions in AGASA.  The value of $ ln\mathcal{R}$ for the actual 94 event dataset with the error assignments used here is 12.88, whereas $ ln\mathcal{R} = 12.98$ with the non-gaussian errors used in \citep{HRGF}.  Third, \citet{HRGF} include the variation of the HiRes exposure density over the sky, whereas here the HiRes exposure is fixed to a constant, 0.2 sr$^{-1}$, which is 1.07 times the average unit-normalized exposure.  This value is chosen to give the correct rate of random 4-fold coincidences, based on approximating the HiRes exposure shown in Fig. 1 of \citep{HRGF} and using the analytic discussion in the previous section to see that the rate of n-fold coincidences $\sim  \int (N_{AG} R_{AG}(\vec{\theta}) + N_{HR} R_{HR}(\vec{\theta}) )^n d^2 \vec{\theta}.
$
When HiRes publishes its exposure in a more accessible form, it will be possible to improve the precision of this estimate.  

\begin{figure}[t]
\epsscale{1.0}
\plotone{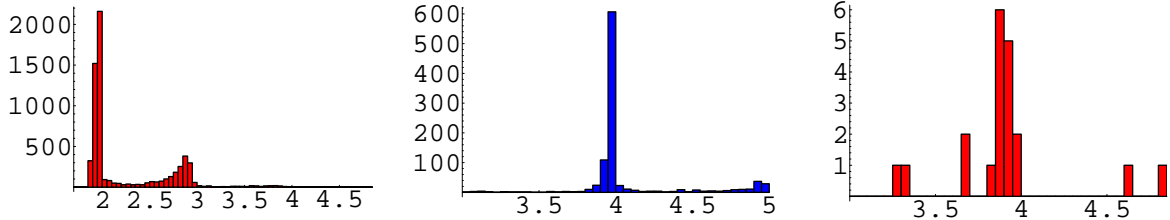}
\caption{Histograms of $n_s$ values as described in the text.
}
\label{nshists}
\end{figure}

Now we turn to the data, and consider different approaches to assessing the likelihood that the 4 events above 37 EeV come from a common source.   The left and right panels of Fig. \ref{nshists} show results from about 17,000 random datasets composed of the 94 events in the combined AGASA and stereo HiRes data above 40 and 30 EeV respectively.  With no condition imposed on $ln \mathcal R$, most trials have $n_s = 1$.  Removing these events by requiring $ln \mathcal R \ge  6$ (left panel) shows that almost all of the remaining trials have $n_s \le 3$; by requiring $ln \mathcal R \ge ln \mathcal R_{\rm obs}$ the background is dramatically reduced as shown in the right panel.  The distribution in $n_s$ for a genuine source is very different from the random case: almost all the trials have $n_s$ greater than $n_{\rm true} - 0.2$.  This is illustrated by the center panel of Fig. \ref{nshists} which shows the $n_s$ distribution for $10^3$ simulated datasets with an embedded quad: 4 events distributed according to the 2-d Gaussians of the quad events, and 90 events distributed at random.  The cut $n_s \ge n_{s, \rm obs} = 3.89$ removes very little of the signal but removes more than half the remaining random background.   
The only condition imposed by \citet{HRGF} was $ln \mathcal R \ge  ln \mathcal R_{\rm obs}$.  Requiring instead that both $ln \mathcal R \ge ln \mathcal R_{\rm obs}$ and $ n_s \ge n_{s,\rm obs}$, the chance probability of the observed quad in about 17,000  datasets is found to actually be $1.0 \pm 0.2 \times 10^{-3}$, where the error is statistical.  

The chance probability is remarkably insensitive to the angular resolution of the events.  To illustrate, consider the magnetic dispersion analysis of the next section where a gaussian random variance $32.5^\circ {\rm EeV}/E$ is added in quadrature to the measurement resolution.   Repeating the simulations above with the new larger errors gives indistinguishable results to within the statistical accuracy used.  While this may seem counter-intuitive, the analytic discussion of the previous section shows it is to be expected at some level of accuracy: differentiating equation (\ref{lnRapprox1}) with respect to $\sigma$, one finds that for the relevant range of $N,~n_s,~\sigma$ and $R$, the change in $ \chi^2$ cancels most of the change in the background probability term.  Of course, some changes in the relative errors of different events can make a significant change in the likelihood, so this insensitivity to resolution is generic buit not universal.  It can also be understood by noting that the chance probability of finding 4 random hits in a clustered configuration depends only on how the spacing of the clustered hits compares to the average density of events, and has no relation whatever to the accuracy of the measuring apparatus, which is solely relevant for determining the quality of the fit to the source hypothesis.  As a result of this natural insensitivity to the resolution, the conclusion that the quad events have a common source is unlikely to change when more precise resolutions become available for individual AGASA events.

The chance probability of 0.1\% is not strictly an {\it a priori} probability because the choice of the energy threshold of the original AGASA dataset used by \citet{AGclus96} was not specified before that analysis, and the correct relative normalization of energies between AGASA and HiRes is uncertain.  However these are minor issues in determining the chance probability of the quad.  Since the energy of the events is not inherently important to the cluster analysis, the threshold is only relevant to the density of background events.  The original AGASA doublet events which are members of the quad have energies 77.6 and 55 EeV -- well above the 40 EeV threshold for the AGASA dataset, and the HiRes member with energy 37.6 EeV is well above the threshold of 30 EeV so there is no issue that the background is being unnaturally minimized by these thresholds.  (Analyses relating to the {\it total number} of doublets or higher multiplets are sensitive to the definition of the dataset, since one of the three original AGASA doublets had a 43 EeV event so that shifting the threshold say to 50 EeV significantly changes the conclusions \citep{finWest}.) 

There is another measure of the probability that the quad is real, which is completely independent of the dataset choice.  One can ask how often {\it starting with the initial AGASA dataset} of 30 events \citep{AGclus96} and adding 27 more AGASA  and 37 HiRes events at random, does one find $ln \mathcal R \ge ln \mathcal R_{\rm obs},~ n_s \ge n_{s,~\rm obs}$.  The result is 32 cases in $5 \times 10^4$ tries, for a chance ``promotion probability" of $0.7 \pm 0.1 \times 10^{-3}$, of which more than half of this comes from the ``C1" AGASA doublet, which was not in fact promoted. 

There are many ``interesting" anomalies which could occur in a random dataset, and the question comes up whether to sum over the probability of getting all possible combinations of $n_2$ doublets, $n_3$ triplets, etc, when computing the random background, which would increase the background estimate.  The correct procedure depends on the theoretical perspective.  If all sources are of comparable apparent luminosity and magnetic smearing is negligible, multiple clusters are expected and their presence or absence can be used to validate or exclude such a model.  By contrast, for the picture inspiring this study -- that different sources probably have substantially different apparent luminosities and cosmic magnetic fields are probably very inhomogeneous, so that few if any sources can be reconstructed -- one has no {\it a priori} expectation to find multiple significant sources and the confidence in the hypothesis is {\it not} increased by having multiple sources.  Rather, the presence or absence of multiple reconstructed sources gives information on the nature of the sources and fields.  In this scenario, the important question is how well the observed cluster fits the hypothesis; the probability that the cluster in question could happen by chance determines how much credence should be attached to its implications while awaiting further data.

\section{Magnetic Deflection and Dispersion}
The strength and direction of the local galactic magnetic field is known \citep{heiles96} but its decay with distance from the local spiral arm and the partition between random and coherent components are quite uncertain. The measured local field may well be dominated by the random component, yet its direction might be correctly modeled, since global information constrains the direction of the coherent field for a given assumed structure.  Therefore, I consider three cases for the galactic magnetic field below: {\bf a)} Taking the coherent magnetic deflection to be negligible, appropriate if the local field is dominated by the random component. {\bf b)} Taking the direction of galactic magnetic deflection to be fixed as in standard models but the magnitude of the galactic deflection, $G_*/E$, to be free.  This is appropriate if the global constraints correctly determine the general direction of the coherent field along the trajectory but, for instance, not how rapidly it decreases away from the disk.  {\bf c)} Taking the direction of deflection as well as $G_*$ to be free.  

In the limit of sufficiently many small deflections, the smearing of cosmic rays from a pointlike continuous source due to random fields is described by a 2-d Gaussian probability distribution, $\frac{1}{2\pi\sigma^{2}} \exp\left(-\frac{(\Delta\vec{\theta})^{2}}
     {2\sigma^{2}}\right) $ of width 
\begin{equation} \label{magsmear}
\sigma^2_B(E) \equiv \left( \frac{E^*}{E} \right)^2= \frac{2 \langle B^2 \lambda \rangle D}{9 E^2} ~~,
\end{equation}
where $\lambda$ is the coherence length of the field, $D$ is the distance to the source, and $\langle B^2 \lambda \rangle$ denotes a weighted mean of $B^2 \lambda$ over the trajectory\citep{waxmanME}.  I assume throughout that all events in the cluster are protons; there are too few events to permit generalizing the analysis to different charges at this time.  Since the experimental angular resolution of each event is also approximately a symmetric 2-d Gaussian, of width denoted $\sigma_{i,0}$,  magnetic smearing increases the effective size of the individual events' resolutions in the fit to $\sigma_i = \sqrt{\sigma_{i,0}^2 + \sigma_B(E_i)^2}$. Barring a caustic, two cosmic rays from the same source with the same charge and energy follow the same trajectory and have no relative dispersion.  Since a caustic only operates over a very narrow range of energy it is {\it a priori} improbable for any given energy, so the {\it relative} magnetic dispersion between the 55 and 53.5 EeV events should be minimal.  Therefore when determining the magnetic smearing we first combine these two events, replacing them by a single event with arrival direction \{$170.22^\circ, ~56.57^\circ$\}, energy $E=54.25$ EeV, and $\sigma = 0.52^\circ$. 

It turns out that cases {\bf a)} and {\bf b)} are practically the same, because when the direction of the coherent deflection is fixed toward $-38.6^\circ$ in local equatorial coordinates, both quad and quint are fit best with $G_* = 0$.  This is also true for case {\bf c)} when all five events are used; it is not meaningful to study this case with just the quad, since there are 6 measured positions and 5 parameters to be constrained, so the uncertainty is too large.  Given the inherent uncertainties, $G_* \approx 0$  is consistent with known constraints on the strength and extent of the local coherent field, as will be developed in greater detail elsewhere.   The magnitude of deflection predicted by standard models gives a poor fit to the observed arrival directions of the quad, and a very poor fit for the quint (G. Medina-Tanco, private communication).

Now the magnetic smearing can be estimated.  The simplest procedure is to adjust $E_*$ to the value which causes $\chi^2$ for the cluster to be equal to the number of degrees of freedom in the fit, as expected on average.  One can also choose the value of $E_*$ which maximizes $ln \mathcal R $; in general this gives a lower value of $E_*$ since smaller $\sigma$ means lower background.  We adopt the former because it is computationally easier but keep in mind that the true value of $E_*$ may be lower.  With $G_* = 0$, there are three free parameters: the coordinates of the source and $E_*$.  The ``quad" consisting of the 77.6, ``54.25", and 37.6 EeV events is therefore described by a $\chi^2$ distribution for 6 - 3 = 3 degrees of freedom, and a first estimate of $E_*$ is that which gives $\chi^2 = 3$ for the ``quad", namely $E_* = 32.5^\circ$EeV. Fortunately, $E_*$ is quite insensitive to a possible systematic offset in the energies between AGASA and HiRes events.  Each experiment has an energy uncertainty of $\sim 15\%$ and a commonly evoked explanation for the discrepant spectra is that AGASA and HiRes energies are systematically 15\% too high and low respectively. (Indeed, awareness of this possibility is why the HiRes threshold of 30 EeV was investigated in \citep{HRGF}.)  But rescaling the AGASA energies by 0.85 and the HiRes energy by 1.15, leads to $E_* = 32.3^\circ$EeV.

If the fifth event does indeed come from the same source, including all 5 events in the analysis will give a more sensitive determination of $E_*$ both because the number of events increases so the $\chi^2$ distribution becomes narrower and because the low energy event is more sensitive to $E_*$.  However to help judge whether the 5th event is likely to be from the same source, we first compute $\chi^2$ for the ``quint" using the value $E_* = 32.5^\circ$EeV obtained from fitting the quad.  The result, $\chi^2  = 3.5$, is a better-than-average fit since there are now 5 degrees of freedom.  The fact that $\chi^2$ for the quint is so good, using the $E_*$ fit to the quad, is evidence that the 5 events are really from the same source and that our treatment of magnetic effects is robust.  If the fifth event had instead been a random background event it most probably would have fallen as far as possible from the quad source location, since that is where the phase space is largest, and the $E_*$ required to get a reasonable fit to the quint would have been larger than found with just the quad.  

Estimating $E_*$ from the ``quint" as the value for which $\chi^2 = 5$, gives $E_* = 23.9^\circ$EeV and a source location of $\{169.40^\circ,~ 56.78^\circ\}$, with a 90\% error radius of $\approx 1^\circ$.  This is the same as that obtained with the quad, $\{169.24^\circ,~ 56.81^\circ\}$, within the uncertainty.   Even though the 5th event gives a better-than-average fit to the point-source hypothesis,  decreasing the HiRes energy threshold to 10 EeV increases the dataset from 94 to 308 events, so there is a significant (1 in 6) chance the fifth event is an accidental association.  The promotion probability from doublet to quint is $0.16\%$ -- small but twice that from doublet to quad -- because in this case 251 instead of 37 HiRes events are added to the original 30 and added 27 AGASA events; the promotion probability is insensitive to the value of $E_*$ taken here to be 23.9$^\circ$EeV.  
 
Assuming the 5 events have a common source, we obtain a ``90\% CL" range for $E_*$ as follows.  For 5 degrees of freedom (continuing to merge the 53.5 and 55 EeV events), 10\% of the time $\chi^2> 9.2$ and 10\% it is $< 1.6$.  The values of $E_*$ giving those values of $\chi^2$ are $E_* = 12^\circ$EeV and $55^\circ$EeV respectively.  The analogous range using just the ``quad" is $(5 - 96)^\circ$EeV, confirming the stronger constraint which is possible when all 5 events are used.   If the energy of the fifth event is lower (higher) than the 15 EeV used above, its effective $\sigma$ would be greater (less) for a given $E_*$, changing the value of $E_*$ derived.  Taking the energy to be 10 EeV, the best value and ``90\% CL" ranges become $20(9 - 52)^\circ$EeV, while for $E_5 = 30$ EeV, $E_* = 30(18 - 65)^\circ$EeV.  Until the energy of the fifth event is made available, we can best characterize $E_*$ as $23.9^\circ$EeV with range $(9-65)^\circ$EeV, always remembering that the maximum likelihood results are somewhat lower.

Equation (\ref{magsmear}), which translates into $E_* = \sqrt{2 \langle B^2 \lambda \rangle D}/3$, is valid only when the net deflection angle times distance to the source is large compared to the typical coherence length of the field: $E_*/E \times D\gtrsim \lambda $.  There is a strong deficit of $L_*$ or brighter galaxies in the direction of the UHECR cluster, out to 140 Mpc (G. R. Farrar, A. A. Berlind and D. W. Hogg, 2005, in preparation). The range of coherence lengths for extragalactic fields usually seen in the literature is $\sim 0.1 - 1$ Mpc, so with $\lambda \equiv \lambda_{\rm Mpc}$ Mpc and $D \equiv 140 ~D_{140}$ Mpc, the condition of validity is $E \lesssim 70~ D_{140}/\lambda_{\rm Mpc}$ EeV.   Our deductions about the magnetic field structure should therefore be trustworthy if the source is beyond the void, particularly since lower energy events are most crucial in determining $E_*$.  However if the source is from one of the unremarkable, low-luminosity galaxies around $D \approx 26$ Mpc, our results are only qualitatively valid unless $\lambda \lesssim 0.2$ Mpc.  Note that an additional criterion must also be met if the Gaussian-random-variable approximation is to be accurate. Namely, the various members of the cluster apart from the merged 53.5 and 55 EeV pair, must sample independent and uncorrelated regions of magnetic field en route to Earth.  This requires $E_*|\frac{1}{E_i} - \frac{1}{E_j}| \gtrsim \lambda /D$, which is more stringent than the condition on the highest energy event alone but may be satisfied at least marginally.  As more events become available, it will be necessary to improve upon the analysis of \citet{waxmanME} and include correlations between the deflections of events of neighboring energies, unless $\lambda/D$ is very small.   
 
Keeping in mind the caveats above, we convert our best value $E_* = 23.9^\circ$EeV to an estimate of the integrated extragalactic fields. (It is qualitatively clear, given the very small scales involved in the random galactic fields compared to the Larmour radius of a proton, $R_L = 108 \frac{ E_{\rm EeV}}{B_{\mu \rm G} }$ kpc, that deflection by random galactic fields is comparatively insignificant as was recently shown in \citet{TTrand}.)  From  eqn (\ref{magsmear}):
\begin{equation} \label{result}
\sqrt{ \langle B^2 \lambda \rangle D} \lesssim  0.96 ~(0.36 - 2.6)~[1.3] ~{\rm nG~Mpc}~~,
\end{equation}
where the uncertainty due to the energy of the 5th event not being available is given in parentheses and the value is square brackets is that derived from the quad alone.   If the source is in fact in a galaxy beyond the void which extends to 140 Mpc in the direction of the cluster, we can obtain an estimated upper limit on the rms field strength in the foreground void:  $\langle B_{\rm void} \rangle ~\lesssim~ 0.07 /\sqrt{\lambda_{\rm Mpc}}$ nG.  This is compatible with Faraday Rotation estimates from distant quasars, as shown by the sheets-voids model of \cite{fp99}.  It is also below the equipartition value, very crudely estimated by taking the void material to be fully ionized, with an electronic density 10\% of the mean baryonic density, and a temperature of $T_3~10^3$K as suggested by \citet{fssh04}, giving $B_{\rm equi} = 0.3 ~T_3^{1/2}$ nG.  

\section{Summary and Conclusions} 
Four cosmic ray events in the combined stereo HiRes and AGASA data above 30 and 40 EeV are consistent with having a common origin, with a chance probability of $10^{-3}$.  When the HiRes threshold is lowered to 10 EeV, the dataset increases from 94 to 308 events and a fifth event consistent with coming from the same source is found.  However due to the large number of added events, there is a 1 in 6 chance that it is background.  The best estimate of the source direction is $\{169.3^\circ,~56.8^\circ\}$ in J2000 equatorial coordinates, with an $\approx 1^\circ$ 90\% error radius.  No coherent net deflection, expected from galactic magnetic fields, is required.  The integrated magnetic smearing is used to infer the order of magnitude constraint $\sqrt{ \langle B^2 \lambda \rangle D} \approx  1 ~{\rm nG~Mpc}$, valid for sufficiently large $D$ or small $\lambda$.  If the source is beyond the void which extends to a distance of 140 Mpc (Farrar et al, 2005, in preparation), we obtain an upper limit on the magnetic field in the void:  $\langle B_{\rm void} \rangle \lesssim 0.07 /\sqrt{\lambda_{\rm Mpc}}$ nG.  With the low statistics available at this time, these constraints on the intervening magnetic fields must be considered qualitative rather than quantitative, but it is interesting that magnetic fields have such small effects in at least one arrival direction. 

The existence of a single source of 4 or 5 events excludes models in which UHECRs are products of the decay or annihilation of superheavy relic particles, unless a mechanism can be found to cause an extreme and concentrated overdensity of relics along this line-of-sight which seems highly unlikely.  This cluster also adds to the difficulties of the $Z$-burst model, because within that model such a cluster could only reasonably occur if produced by the same (tremendously powerful) source of neutrinos, presumably at a cosmological distance.  But in that case, there should be negligible magnetic smearing within the cluster since the observed UHECRs would be mainly photons from $\pi^0$'s produced in a local neutrino DM halo.

A probability of $\approx 0.1\%$ that the quad is due to a statistical fluctuation corresponds to its being a ``3.3 $\sigma$" effect.  This is well above threshold for being taken seriously but is not definitive.  More data will be very helpful -- higher energy events to establish the signal more firmly (or debunk it) and lower energy events to permit a more detailed study of the source and magnetic field. HiRes is continuing to run and AGASA plans to release its lower energy data soon, however these data will make a fairly meager change in the statistics.  The tremendous physics payoff of having much better statistics on this cluster provides additional motivation for a new northern hemisphere UHECR detector with large aperture and good resolution since it is not in the field of view of the Pierre Auger Southern Observatory.

\acknowledgments
I particularly wish to thank Stefan Westerhoff and Chad Finley, whose ideas, criticisms and questions have been invaluable, and I am indebted to M. Teshima and the HiRes collaboration for information about their experiments. I have also benefited from discussions with and input from S. Furlanetto, C. Heiles, G. Medina-Tanco, A. Mincer, G. Sigl, and V. Springel.  This research has been supported in part by NSF-PHY-0101738 and NASA NAG5-9246; simulations were performed on a linux cluster acquired through NSF-MRI-0116590.


\end{document}